\documentclass[A4paper]{PoS}
\usepackage{aas_macros}
\usepackage{graphicx}
\usepackage{amsmath}
\usepackage{amssymb}
\usepackage{setspace}
\usepackage[utf8x]{inputenc}
\usepackage{latexsym}
\usepackage{color}
\usepackage{dcolumn}
\usepackage{bm}
\title{Pop III GRBs: an estimative of  the event rate for future surveys}

\ShortTitle{Pop III GRBs forecast}

\author{\speaker{Rafael S. de Souza}\\
        Author affiliation\\Korea Astronomy \& Space Science Institute, Daejeon 305-348, Korea\\
Max-Planck-Institut f\"ur Astrophysik, Karl-Schwarzschild-Str. 1, D-85748 Garching, Germany\\
        E-mail: \email{rafael@kasi.re.kr}}

\abstract{We discuss the  theoretical event rate of gamma-ray bursts (GRBs) from the collapse of massive primordial stars.  We construct a  theoretical model to calculate the rate and detectability of these GRBs taking into account all important feedback and  recent results  from numerical simulations of pristine gas.  We expect to observe a maximum of  N $\lesssim$ 0.2 GRBs per year integrated over at z > 6 with \textit{Swift} and N $\lesssim$ 10 GRBs per year integrated over at z > 6 with  EXIST
 (assuming   a sensitivity  up to 10$\times$  higher than \textit{Swift}). }

\FullConference{Gamma-Ray Bursts 2012 Conference -GRB2012,\\
		May 07-11, 2012\\
		Munich, Germany}

\begin{document}

\section{Introduction}

The first stars (Pop III) in the Universe  are likely to   have 
played a major  role in the early cosmic evolution,  
by emitting the first light and producing the first heavy elements
\cite{Yoshida2008,Bromm2011}.
It is of great importance to understand the origin and evolution of such objects, since their 
detection would permit to probe the pristine regions of the Universe.

Pop III stars may produce 
collapsar gamma-ray bursts (GRBs) whose total isotropic energy 
could be $\approx 2$ orders of magnitude larger than average
\cite{komissarov2010,meszaros2010,toma2011}. 
Even if a Pop III star has a supergiant
hydrogen envelope, 
the GRB jet can break out of it due to the long-lasting 
accretion of the envelope itself \cite{suwa2011,nagakura2011}.

\section{GRB redshift distribution}
The following is a brief overview of the model we used. Readers are encouraged to check \cite{rafael2011} for a full description.
As  long GRBs  are expected to follow the death of  very massive  stars,  their rate  provide an useful probe for  cosmic star formation history \cite{rafael2011,ishida2011, rafael2011b,robertson2012}. 
 
 Over a particular time interval,  $\Delta t_{\rm obs}$,  in the observer rest frame, 
the number of observed GRBs originating between redshifts $z$ and $z + dz$
is
\begin{equation}
\frac{{\rm d}N_{\rm GRB}^{obs}}{{\rm d}z} =\frac{\Omega_{obs}}{4\pi}\eta_{beam} \eta_{\rm GRB}\Psi(z)\frac{\Delta t_{\rm obs}}{1+z}
\frac{{\rm d}V}{{\rm d}z}\int_{\log{L_{lim}(z)} }^{\infty}p(L)d\log{L},
\label{dngrbtrue}
\end{equation}
where  $\eta_{\rm GRB}$ is the GRB formation efficiency,   $\Psi$ is the cosmic star formation rate (SFR) density,  ${\rm d}V/{\rm d}z$ is the comoving volume element per  redshift unit, $p(L)$ is the luminosity function, $\eta_{beam}$ is the beaming factor of the burst ($\sim 6^{\circ}$) and $\Omega_{obs}$ is the field of view of the experiment.  The adopted values of $\Omega_{obs}$ are 1.4, 2, 4, and 5 for Swift, SVOM, JANUS, and EXIST, respectively.  The intrinsic GRB rate  (the total number of GRBs per year on the sky ) is defined as  $\frac{{\rm d}N_{\rm GRB}}{{\rm d}z}= \eta_{\rm GRB}\Psi(z)\frac{\Delta t_{\rm obs}}{1+z}
\frac{{\rm d}V}{{\rm d}z}$.  

 \subsection{Star Formation History}
 
In order to determine the star formation rate (SFR) at early epochs, we assume  that stars are formed in collapsed dark matter halos.  In what follows, we adopt the Sheth-Tormen mass function  to estimate  the number of dark matter halos, $n_{ST}(M, z)$, with mass less than M per comoving volume at a given redshift.
The collapsed fraction of mass, $F_{\rm col}(z)$, available for Pop III star formation is given by

\begin{equation}
F_{\rm col} = \frac{1}{\rho_{m}}\int_{M_{H_2}}^{\infty} dMMn_{sT}(M,z). 
\label{fcol}
\end{equation}
Where $\rho_m$ is the total mass density of the background Universe. 
The star formation efficiency in the early Universe largely depends on the ability of a primordial gas to cool and condense. Hydrogen molecules ($H_2$) are the primary coolant in a gas in small mass "minihalos", and can be dissociated by  soft ultraviolet radiation. Thus,  a ultraviolet background in the Lyman-Werner (LW) bands can easily suppress the star formation inside mini-halos. We model the dissociation effect by setting the lower limit of equation (\ref{fcol}) as the minimum halo mass capable of cooling by molecular hydrogen in the presence of a Lyman-Werner (LW) background (see \cite{rafael2011} for more details).  

\begin{figure}
\includegraphics[width=0.9\columnwidth]{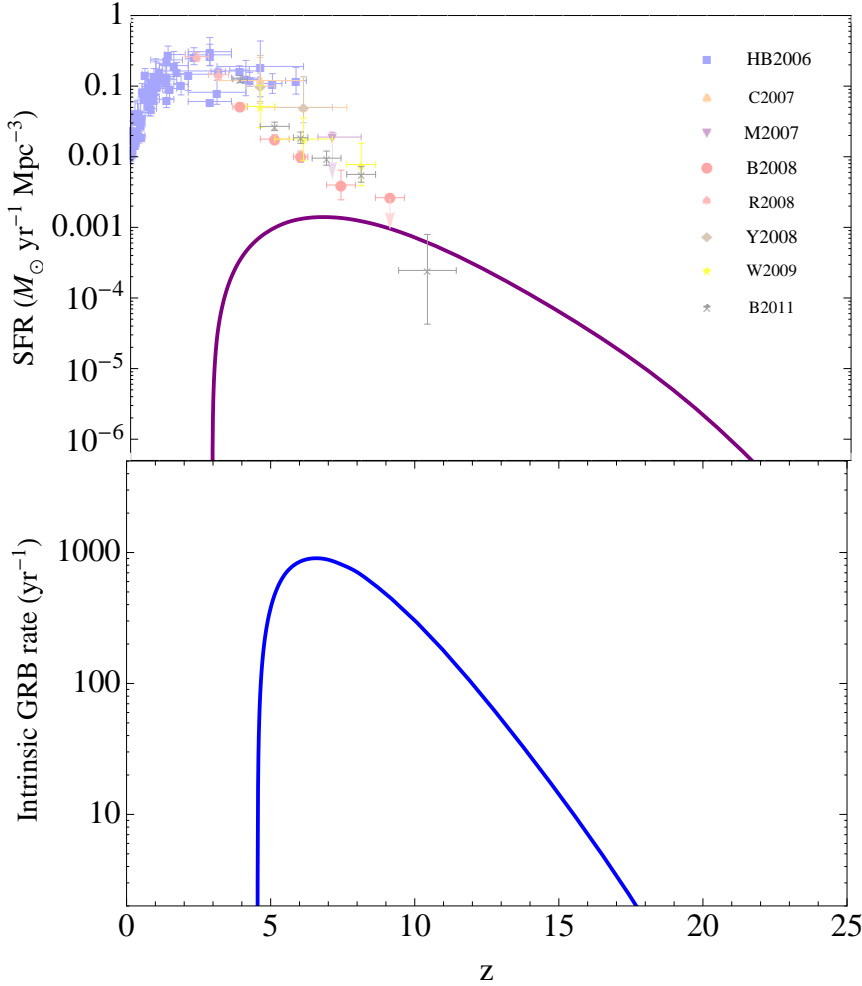}
\caption
{Top: Optimistic model for Pop~III star formation rate. The light points are  independent SFR determinations compiled from the literature. }
{Bottom: The intrinsic GRB rate  ${\rm d}N_{\rm GRB}/{\rm d}z$. In other words, the number of  GRBs per year on the sky
 (on-axis + off-axis).}
\label{fig:SFRIII}
\end{figure}

The star formation density can be described by the following expression, 
\begin{equation}
\Psi(z)= Q_{\rm H II}f_{\zeta}f_{*}\frac{\Omega_b}{\Omega_m}\frac{\rm {d}F_{\rm {col}}}{dt}, 
\label{SFRH}
\end{equation}
where $\Omega_{\rm m}$  and $\Omega_{b}$ are the matter and baryon  cosmological density parameters respectively.  The factors $f_{\zeta}$ and  Q$_{\rm H II}$   represent  the global filling fraction of metals via galactic winds and  the volume filling fraction of ionized regions as shortly described below.  The star formation efficiency  ranges from  $f_{*} = 0.001$ as a conservative choice \cite{Greif2006} to  $f_{*} = 0.1$ \cite{bromm2006}  as an upper limit.  Given the large uncertainties on the above quantities we derive upper limits on the intrinsic GRB rate
 using $\sim1$ year timescale radio variability surveys, which sets upper limits on the number of transients sources in the sky.  Such limits can be used to ruled out  our optimistic case. This is done by relating the intrinsic GRBs rate derived  by our model with the number of afterglows in the sky, which is limited by observational limits from radio surveys (we are not really discussing how the constraints are derived here due to space limitation, please see \cite{rafael2011} for details).  
\textbf{Reionization:}
 Inside growing {H\sc{II}}  regions, the gas is highly ionized and 
the temperature is $\sim 10^4$ K. The volume filling factor of ionized regions, $Q_{\rm {H\sc{II}}}(z)$,   influences   the typical mass of  collapsed stars. 
\textbf{ Metal Enrichment :}
The  metal-enrichment in the inter-galactic medium (IGM)
determines when the formation of primordial  
stars is terminated (locally) and 
switches from the Pop~III mode to a more conventional mode of star formation.  
We  follow the metal-enriched wind propagation outward from a central galaxy in order to  evaluate the average metallicity over cosmic scales as a function of redshift. 
 The top panel of Fig.  \ref{fig:SFRIII} shows  the  upper limit  for   Pop~III  SFR adopted here.   
 
\subsection{Initial Mass Function and GRB Formation Efficiency}

The stellar initial mass function (IMF) is critically
important to determine the Pop~III GRB rate.  The IMF  determines the fraction of stars with minimum mass which is  able to trigger 
GRBs.  
The GRB formation efficiency factor per stellar mass can be written as  
\begin{equation}
\eta_{\rm GRB} =  f_{\rm GRB} \int_{M_{\rm GRB}}^{M_{\rm up}}\phi(m)dm/
\int_{M_{\rm low}}^{M_{\rm up}}m\phi(m)dm,
\end{equation} 
The  $f_{\rm GRB}$  factor gives the fraction of stars in this range of mass that will produce GRBs. 
And  $\phi(m)$ is the stellar IMF which, we considered a Gaussian. 
We assume $\bar{M} = 40 M_{\odot}$  
and  dispersion $\sigma_c =  (\bar{M}-M_{\rm low})/3$.  
$M_{\rm low}$ and  $M_{\rm up}$ are  the minimum  and maximum mass 
adopted for  Pop III stars. Which ranges from   
$10 M_{\odot}$ to $100 M_{\odot}$ respectively.  
$M_{\rm GRB}$ is the minimum mass  necessary to trigger 
GRBs, which we set to be $25 M_{\odot}$ \cite{bromm2006}.

\subsection{Luminosity function}

The number of GRBs detectable by any given instrument depends on the 
instrument-specific flux sensitivity threshold and also on the intrinsic 
isotropic luminosity function of GRBs. We adopt a power-law distribution function  similar to 
 \cite{wanderman2010}


\begin{equation}
\label{LF}
p(L) =\left(\frac{L}{L_{*}}\right)^{\alpha} 
L<L_{*}, ~
\left(\frac{L}{L_{*}}\right)^{\beta} 
L >L_{*}, \\
\end{equation}
where $L_{*}$ is the characteristic isotropic  luminosity, $\alpha = -0.2^{+0.2}_{-0.1}$ and $\beta = -1.4^{+0.3}_{-0.6}$. 
The Pop III GRBs are assumed to be energetic with isotropic 
kinetic energy $E_{iso}\sim 10^{56-57} erg$ but long-lived,  $T_{90} \sim 1000~ s$. Thus
 the luminosity would be moderate,   
$L_{*}\sim\epsilon_{\gamma}\times 10^{56-57}/1000\sim10^{52-53}$ ergs/s 
if $\epsilon_{\gamma}\sim 0.1$ is the conversion efficiency from the jet kinetic 
energy to gamma rays  \cite{suwa2011}.

Using equation (\ref{LF})  we can predict the observable GRB rate 
for the \textit{Swift},  SVOM, JANUS,  and EXIST missions.   
For \textit{Swift}, we set a bolometric energy flux limit 
$F_{\rm lim} = 1.2 \times 10^{-8} {\rm erg}~ {\rm cm}^{-2}~ {\rm s}^{-1}$. We adopt a similar limit for SVOM.  
For JANUS, $F_{\rm lim} \sim 10^{-8} {\rm erg}~ {\rm cm}^{-2}~ {\rm s}^{-1}$.   
The luminosity threshold is then $L_{\rm lim} = 4\pi\, d_{\rm L}^{2}\, F_{\rm lim}$.
Here $d_L$ is the luminosity distance for the adopted $\Lambda$CDM cosmology. 
Given that EXIST is expected to be $\sim 7-10$ times more sensitive 
than \textit{Swift},  
we set its sensitivity threshold ten times lower  
than \textit{Swift}'s as an approximate estimative. 
For simplicity,  we assume that the spectral energy distribution (SED) 
peaks at X-to-$\gamma$ ray energy (detector bandwidth) as an optimistic case.   
\begin{figure}
\includegraphics[width=0.9\columnwidth]{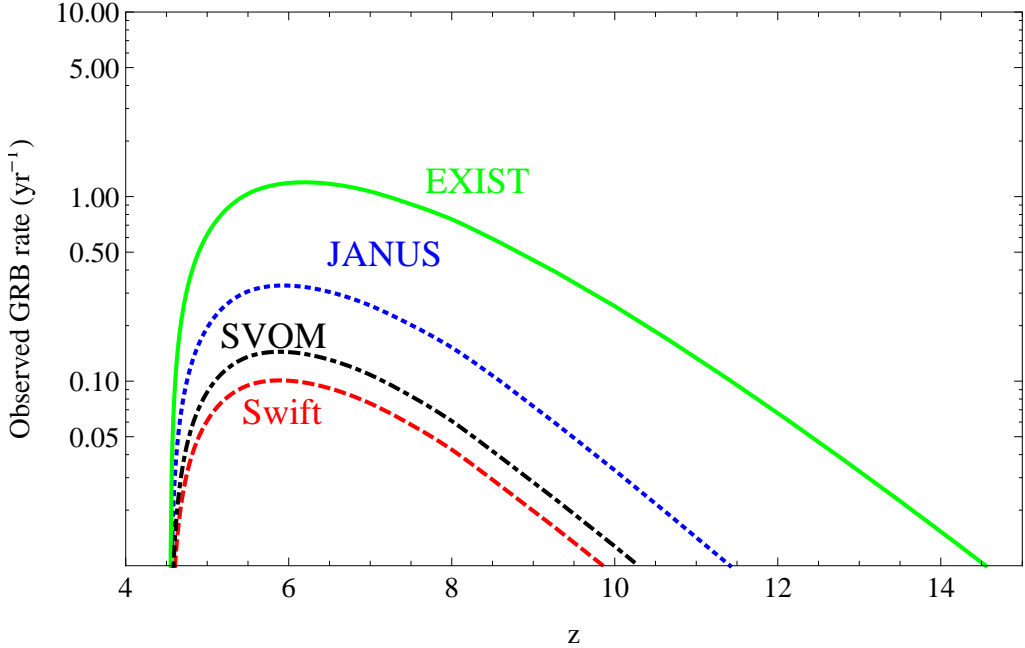}
\caption{
Predicted Pop~III observed GRB rate, ${\rm d}N_{\rm GRB}^{obs}/{\rm d}z$. Those observed by \textit{Swift}, dashed red line;   SVOM, dot-dashed black line;
JANUS, dotted blue line;  and EXIST, green line;  for our model with
Gaussian  IMF, $f_{*}=0.03$, $f_{GRB} = 0.001$.}
\label{fig:GRB3}
\end{figure}
We expect  to observe a maximum of N $\lesssim$ 10 GRBs per year integrated over at z > 6 with EXIST, and N $\lesssim$ 0.2  GRBs per year integrated over at z > 6 with Swift.

\section{Summary}

GRBs can be observed at large distances, been a powerful tool to probe the early Universe \textit{in situ}. Together with pair instability supernovae, they are probably the most promising way to observe the first stars. Here we  discuss the observed rate of such objects in current and future surveys. We take into account the recent developments concerning the origin of Pop III stars, in particular the last results from \cite{hosokawa2011}, which show Pop III stars can be less massive than previously thought.  
We expect  to observe a maximum of N $\lesssim$ 10 GRBs per year integrated over at z > 6 with EXIST, and N $\lesssim0.2$  GRBs per year integrated over at z > 6 with Swift. If we consider  that we have no observation so far from Pop III GRBs among all GRBs observed by \emph{Swift} with redshift measurements. We can place an upper limit over the fraction of Pop III GRBs over Pop II/I ones, $< 1/N_{Swift}$, where $N_{Swift}$ is the number of long GRBs with redshift measurements ($\sim 150$). Which gives an fraction of $\sim 0.0067$. Thus , considering an average of 150/6 GRBs $yrs^{-1}$ (25 GRBs $yrs^{-1}$ ) with redshift measurements.  It gives un upper limit of $N \lesssim 0.56 $ Pop III GRBs $yr^{-1}$ ($0.0067\times 25\times1/0.3$), taking into account that we only measure redshifts for $\sim 0.3$ of all GRBs. Which is consistent with our estimative. 

It's important to note that even if such objects are above the flux limit threshold,  most of the  high-z GRBs might be hidden by dust and be "dark grbs". An alternative is  to detect their afterglows. As their afterglows are expected to be long lasting $\sim 1000$ days, will probably be difficult to be detected  them as transients. To overcome this problem,  we should rely in surveys strategies  that take snapshots from the  sky during long and spaced  periods of time. Several tools for photometric supernova classification already exist and have been tested with synthetic supernovae  light curves \cite{ishida2012}. We suggest  that  the same approach could be done in order to look for Pop III GRBs candidates in radio surveys.

\end{document}